\documentclass[sigconf,screen,nonacm]{acmart}
\usepackage{booktabs}
\usepackage{comment}
\newcommand{\ptitle}[1]{\textbf{#1}.}

\AtBeginDocument{%
  }

\usepackage{xcolor}

\begin{document}

\title{Are LLMs ready to help non-expert users to make charts of official statistics data?}

\author{Gadir Suleymanli}
\affiliation{%
  \institution{Ghent University}
  \country{Belgium}
}

\author{Alexander Rogiers}
\affiliation{%
  \institution{Ghent University}
  \country{Belgium}
}

\author{Lucas Lageweg}
\affiliation{%
  \institution{University of Amsterdam \& CBS (Statistics Netherlands)}
  \country{Netherlands}
}

\author{Jefrey Lijffijt}
\affiliation{%
  \institution{Ghent University}
  \country{Belgium}
}

\renewcommand{\shortauthors}{Suleymanli et al.}

\begin{abstract}
	In this time when biased information, deep fakes, and propaganda proliferate, the accessibility of reliable data sources is more important than ever. National statistical institutes provide curated data that contain quantitative information on a wide range of topics. However, that information is typically spread across many tables and the plain numbers may be arduous to process. Hence, this open data may be practically inaccessible. We ask the question \emph{``Are current Generative AI models capable of facilitating the identification of the right data and the fully-automatic creation of charts to provide information in visual form, corresponding to user queries?''}.
	
	We present a structured evaluation of recent large language models' (LLMs) capabilities to generate charts from complex data in response to user queries. Working with diverse public data from Statistics Netherlands, we assessed multiple LLMs on their ability to identify relevant data tables, perform necessary manipulations, and generate appropriate visualizations autonomously. We propose a new evaluation framework spanning three dimensions: data retrieval \& pre-processing, code quality, and visual representation. Results indicate that locating and processing the correct data represents the most significant challenge. Additionally, LLMs rarely implement visualization best practices without explicit guidance. When supplemented with information about effective chart design, models showed marked improvement in representation scores. Furthermore, an agentic approach with iterative self-evaluation led to excellent performance across all evaluation dimensions. These findings suggest that LLMs' effectiveness for automated chart generation can be enhanced through appropriate scaffolding and feedback mechanisms, and that systems can already reach the necessary accuracy across the three evaluation dimensions.
\end{abstract}





\begin{teaserfigure}
    \begin{figure}[H]
    \centering \includegraphics[width=\textwidth]{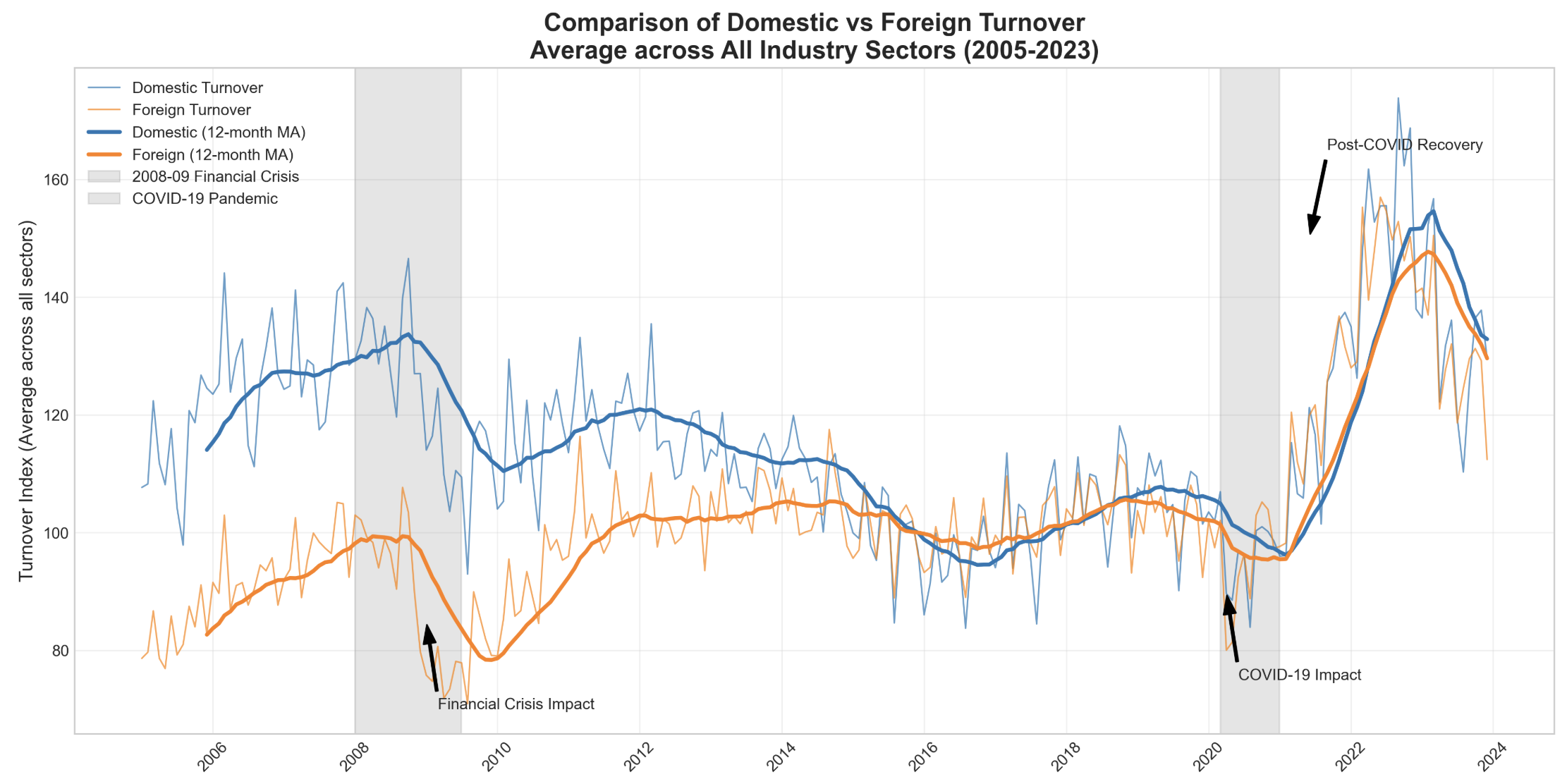} \caption{Sample visualization generated by Claude 3.7 using a modular prompt and a self-feedback loop (7 iterations), responding to the prompt: \textit{``Plot and compare foreign and domestic turnover for All Sectors.''} The chart compares domestic and foreign turnover indices from 2005 to 2023, averaged across all industry sectors, based on data from Statistics Netherlands. Raw monthly trends are shown with thin lines, while 12-month moving averages are overlaid with thicker curves to highlight longer-term patterns. Two key economic disruptions (the 2008–2009 Financial Crisis and the COVID-19 Pandemic) are marked with shaded regions. The visualization showcases use of reasoning, clear axis labeling, and appropriate use of smoothing and context. Sample figures are displayed as produced by the LLM system, without any post-processing. More results can be found at \url{https://github.com/aida-ugent/llm_visualization_results}\label{fig:claude_viz}}
    \end{figure}
\end{teaserfigure}


\maketitle

\section{Introduction}

\ptitle{Official statistics data are more relevant than ever, but are inaccessible}
In an era increasingly characterized by the proliferation of misinformation, ensuring public access to and understanding of reliable official statistics is becoming more and more important \cite{OECD2024}. 
National statistical institutes, such as Statistics Netherlands (CBS), serve as official providers of such authoritative datasets, which are vital for informed public discourse and data-driven decision-making by governments and citizens alike. 
However, these official statistics datasets are often large-scale, complex, and laden with domain-specific terminology, posing significant barriers to direct interpretation, especially for non-expert users.

\ptitle{On-demand retrieval and chart generation using GenAI may improve accessibility}
Data visualization is a powerful means to distill insights from complex information, yet creating effective charts traditionally demands considerable technical expertise and time. 
This limits the ability of many individuals to independently explore and understand official data. 
The recent rise in capabilities of Large Language Models (LLMs) presents an opportunity: their capacity to translate natural language requests into executable code, including visualization scripts, promises to democratize data analysis. 
By enabling users to interact with the official data in plain language, LLMs could significantly lower the thresholds to data-driven decision-making and improve data literacy overall.

\ptitle{GenAI systems may face several challenges and it is largely unknown how well they may tackle these}
Despite this promise, the true readiness of LLMs to assist non-expert users in generating meaningful and accurate charts from the highly structured, real-world data typical of official statistics remains an open question. 
Such a task is non-trivial; an LLM must not only (i) correctly interpret a user's query, potentially expressed using everyday language rather than precise domain jargon, but also (ii) identify the correct data table from potentially thousands, (iii) perform necessary data manipulations like filtering and aggregation accurately, (iv) select an appropriate chart type for the data and query, and (v) generate syntactically correct code that adheres to visualization best practices. 
An error at any stage can lead to flawed or misleading visualizations, undermining the goal of clear communication. 
If these challenges can be overcome, LLMs could enable a broad audience, from researchers and journalists to the general public, to obtain reliable visuals from authoritative data sources in seconds, thereby widening access to official statistics.

\ptitle{Contributions}
To explore the current possibilities, we designed 25 questions of various difficulty that may be answered using data from seven frequently-accessed tables from Statistics Netherlands, and tested the capabilities of eight state-of-the-art LLMs of varying sizes to generate corresponding charts that would answer these questions.
We created a comprehensive evaluation framework to score each attempt by the models on 22 unambiguous binary questions---covering visual clarity, data correctness, and code soundness---yielding quantitative, model-agnostic metrics. These questions are specifically selected to not only base the evaluation solely on the generated design but also the quality of code and relevance.
Through this study, we aim to answer the question to what degree is it feasible (with current tools) to incorporate said models in a data visualization pipeline for a real-world use case.

In summary, this research delivers three contributions: (1) an agentic system that combines targeted prompt engineering with iterative self-correction for natural-language chart generation on official statistics; (2) a comprehensive evaluation framework for assessing the model performance; and (3) a comparative study showing how model size, provided context, and task difficulty each influence end-to-end visualization quality.

The evaluation results suggest that, when supplied with a good scaffold such as compact design guidance and the freedom to debug their own outputs, modern LLMs can already produce highly effective charts for real-world queries, opening a path toward accessible, language-driven analytics for non-experts.

\ptitle{Outline} We first review related work on automated data visualization and LLM applications in Section~\ref{sec:relatedwork}. In Section~\ref{sec:systemdesign} we present the system design, including prompt design and the implementation of both zero-shot and agentic LLM systems. In Section~\ref{sec:evaluation}, we detail the structured evaluation framework with criteria across visual, data, and code dimensions. Results are presented in Section~\ref{sec:results}, and we discuss key findings and conclusions in Section~\ref{sec:discussion}, with implications for future research and real-world deployment.


\section{Related Work\label{sec:relatedwork}}
Recent advances in large language models (LLMs) have expanded their applications into areas traditionally requiring substantial domain expertise, including data visualization. Conventional tools such as D3.js, Vega-Lite, Plotly Express, and Tableau provide powerful capabilities for visual representation but effective use of these tools often demands familiarity with programming concepts, chart design principles, and domain-specific data structures.

In response, research interest in automating the visualization process has grown rapidly, aiming to lower the technical barriers to data exploration and communication. The Machine Learning for Visualization survey \cite{ieee2022ml4vis} highlights a notable surge in visualization-related studies, particularly since the widespread adoption of LLMs. The potential to translate natural language queries directly into high-quality visual outputs offers an exciting opportunity to make data analysis accessible to a broader audience. Yet, De Bie et al. \cite{debie2022automating} argue that, in the context of data science, the parts of the workflow outside of pure model construction, such as data preprocessing, exploratory analysis, results communication remain especially hard to automate due to their open-ended, domain-specific nature that is reliant on human judgement. So, achieving reliable and precise visualization generation remains a significant technical challenge. 

To address these challenges, a variety of toolkits and systems have been developed, leveraging natural language processing and machine learning techniques. NL4DV converts English queries into JSON analytic specifications and corresponding Vega-Lite charts \cite{nl4dv}. ChartGPT extends this idea by fine-tuning FLAN-T5-XL on a six-stage reasoning pipeline, tackling vague or incomplete requests through step-wise code generation \cite{chartgpt}. Recommendation engines such as DeepEye \cite{deepeye} and VizML \cite{vizml} predict suitable chart encodings from data characteristics, while LLM4Vis employs few-shot prompting with ChatGPT to suggest multiple alternatives together with textual justifications \cite{llm4vis}. Integrated frameworks then began to appear: LIDA orchestrates four modules, namely summarizer, goal explorer, visualization generator, and infographer, around GPT-3.5 to move from raw table to full infographic in a single workflow \cite{lida}. At the fully automated end, Data2Vis maps tabular inputs to Vega-Lite specs with a sequence-to-sequence recurrent network, eliminating the need for handcrafted templates \cite{data2vis}.

The arrival of stronger foundation models has encouraged lighter, prompt-centric designs. CHAT2VIS demonstrates that carefully engineered prompts alone can push GPT-3/Codex to produce runnable matplotlib scripts and even handle multilingual or ambiguous utterances \cite{chat2vis}. Prompt4Vis further boosts zero-shot performance by automatically mining in-context examples and pruning irrelevant schema tokens before each LLM call, reducing the gap with bespoke neural architectures \cite{prompt4vis}.
MatPlotAgent is an agentic framework that writes MatPlotLib code and was evaluated for scientific data visualization \cite{matplotagent}. Finally, VisPath refines the agentic approach of MatPlotAgent to enhance performance, as evaluated on two code-focused benchmarks \cite{vispath}. Although these studies bear similarity to ours, we rely fully on manual evaluation of the results using a comprehensive set of binary evaluation questions and our prompts are much less specific and more in line with use by non-experts. 

Complementing these heavyweight engines, retrieval-augmented solutions such as VIST5 combine a compact, fine-tuned T5-base model with an external memory of few-shot examples, yielding domain-portable agents that run in real time on commodity hardware \cite{vist5}. In parallel, Visistant combines Google’s Gemini-Pro with LangChain conversational memory, illustrating how modern chat frameworks can support iterative refinement—users can zoom, re-scale, or alter encodings with follow-up instructions \cite{visistant}. Together, these studies chart a spectrum of design trade-offs between accuracy, latency, and resource footprint.


Nevertheless, despite the promising trajectory of these systems, critical challenges remain regarding the accuracy, reliability, and adaptability of LLM-driven visualization generation. Recent evaluations \cite{ready} reveal that although models such as GPT-4 achieve considerable success (it can render roughly 80\% of proposed charts) in generating visualizations using libraries like matplotlib, Seaborn, and Plotly, they are still prone to a range of errors. These shortcomings become particularly evident when dealing with complex datasets or ambiguous user queries, where context understanding and precise mapping are required. Furthermore, current evaluation frameworks are often limited in scope. Automated tools like VisEval \cite{viseval} and VizLinter \cite{vizlinter} have made progress in structuring evaluation around dimensions such as validity, legality, and readability. However, they typically assess outputs based on syntactic correctness rather than semantic fidelity or practical use, overlooking deeper aspects like effective communication, insightfulness, and task alignment.

A key limitation in the existing body of research is its focus on general-purpose datasets, often curated for academic or experimental purposes. The application of LLMs to official, structured data remains a relatively unexplored frontier. Official statistical datasets, such as those maintained by national statistical organizations, present particular challenges due to their complexity, size, and domain-specific characteristics. These datasets frequently involve hierarchical structures, specialized terminologies, and metadata that must be carefully interpreted for effective visualization.

Recent efforts at the Statistics Netherlands have highlighted these challenges. Kouwenhoven et al.\ \cite{kouwenhoven2024constrained} propose a pipeline for statistical question answering that leverages knowledge graphs and constrained decoding techniques to improve retrieval accuracy over large, structured datasets. Their approach significantly enhances query relevance but focuses mainly on text-based answers, without extending into the domain of visualization. In their research, Lageweg et al.\ \cite{lageweg2024generative} utilize schema-constrained decoding to ensure LLM outputs faithful to the data source, improving precision and relevance in statistical contexts. Although both systems demonstrate effective strategies for structured data retrieval, they stop short of integrating automatic visualization as a continuation of the user query process, leaving an important gap in the practical usability of these systems for exploratory data analysis.

In light of these developments, there is a growing recognition that simply generating static visualizations is insufficient. True automation requires systems that can retrieve relevant structured data, interpret complex queries, reason about suitable representations, and produce effective visual outputs within a unified framework. Building on earlier work in knowledge-constrained retrieval and iterative refinement, this paper addresses this need by proposing an agentic framework that enables language model agents to autonomously retrieve, process, and visualize official datasets based on natural language prompts. Through iterative feedback loops, modular prompt engineering, and structured evaluation, the framework aims to overcome limitations identified in existing systems.

\section{System Design\label{sec:systemdesign}}
This section introduces the pipeline that we design to evaluate LLMs capabilities for the visualization tasks. We describe the dataset retrieval process, the design of both zero-shot and agentic systems, the structure of prompts and modular instructions, and the architecture of the tool-equipped agent system

\subsection{CBS Data}
This study uses official statistics published by Statistics Netherlands (CBS), which provides structured datasets on a wide range of topics such as economy, demography, and industry. These datasets are made available through the CBS OData3 API, which also offers English-language access to metadata and raw tabular data in a machine-readable JSON format. Each dataset is organized into a table with a unique identifier, and includes structured metadata describing variables, dimensions, and data types. For instance, the table \texttt{85332ENG}, which contains data on births in Caribbean Netherlands, can be accessed using the following OData query:
\url{https://opendata.cbs.nl/ODataApi/OData/85332ENG/TypedDataSet}.

\subsection{Dataset Retrieval via Sentence Transformers}

For the zero-shot approaches, the first step in the visualization pipeline involves identifying the most relevant dataset based on the user’s natural language prompt. We leverage metadata made available through the OData3 API, which provides descriptions of thousands of structured data tables. These descriptions serve as the reference text corpus for retrieval.

To match user prompts with appropriate data tables, we employ Sentence Transformers—a family of Transformer-based models capable of generating semantically meaningful embeddings for sentences. After testing multiple candidate models (e.g., \texttt{paraphrase-} \texttt{MiniLM-L3-v2}, \texttt{multi-qa-MiniLM-L6-cos-v1}), we selected \texttt{all-} \texttt{mpnet-base-v2} due to its superior performance in correctly handling ambiguous or domain-specific terms. For example, it reliably distinguishes between ``labour'' in the context of employment statistics versus childbirth in fertility data, where smaller models failed.

The retrieval process involves: (1) pre-encoding each metadata description into a dense vector using the selected model and storing them for efficiency, (2) encoding user prompts at runtime with the same model, and (3) computing cosine similarity between the prompt and all stored embeddings. The dataset with the highest similarity score is selected as the most relevant match (top-1 policy; no human filtering). This embedding-based approach enables robust semantic matching beyond keyword overlap, effectively capturing synonyms and contextual meaning.

To assess retrieval effectiveness, we performed an exact match analysis across the 25 evaluation questions. The correct dataset appeared in the top 10 results for 80\% of questions, in the top 5 for 68\%, and ranked first in 36\%. While this indicates that there is room for improvement, it also reflects the challenge of a highly overlapping corpus where many datasets may appear or be equally relevant. We found that most top-ranked results are at least good alternatives, containing either the requested or similar data. In all experiments below, we use the top-1 dataset returned by the retriever. A detailed breakdown of retrieval performance can be found at \url{https://github.com/aida-ugent/llm_visualization_results}.

\subsection{Study Design and Evolution}

Using the top-1 retrieved dataset (no manual filtering), our research followed a two-phase approach to comprehensively evaluate LLM capabilities for visualization tasks:

\ptitle{Phase 1: Zero-Shot Evaluation} We first conducted extensive experiments where LLMs were provided with dataset context and instructions, requiring them to generate complete Python visualization solutions in a single response without iteration or feedback.
    
\ptitle{Phase 2: Agentic System Development} Building on insights from Phase 1, we designed and implemented an agentic system that enabled LLMs to interact with data and code execution environments, enabling an iterative visualization development process.

This natural progression in our research methodology mirrors the evolution of LLM capabilities themselves, from constrained text generation to more sophisticated reasoning and agentic behaviors capable of solving complex tasks through iteration and feedback.

\subsection{System Prompts and Input Engineering}

Prompt design was a critical component in the research methodology, with distinct approaches for zero-shot and agentic systems:
\begin{enumerate}
    \item \textbf{Zero-Shot System Prompt Design}: For our initial experiments, we designed a structured prompt template (Figure \ref{fig:zero-shot-prompt}) focused on maximizing the LLM's ability to generate complete visualization code in one pass. The prompt structure evolved through several iterations, each addressing the limitations observed in earlier outputs (based on early example prompts, not using the 25 final evaluation prompts). Early versions of the prompt provided all of raw data directly to the model, which proved ineffective due to confusion about data types and subset selection. So, raw data was removed, and key components such as column names, sample row, and metadata descriptions were added to clarify the dataset structure without overwhelming the models. Additional clarifications regarding hardcoding and assumptions about the used \texttt{df} variable were added to ensure the model used the full dataset, guiding it to generate correct and generalizable code.
    As a result, the final prompt structure provided comprehensive context about the dataset while establishing clear constraints around code generation.
    
    \item \textbf{Agentic System Prompt Design}: For the agentic system, we developed a more sophisticated prompt (Figure \ref{fig:agentic-prompt}) that guided the LLM through an interactive workflow. 
    This prompt introduced several advanced elements:
    \begin{itemize}
        \item Specific tool definitions for environment interaction
        \item Structured workflow guidance through a seven-step process
        \item Explicit file paths and operational parameters
        \item Error handling and debugging instructions
        \item Verification and feedback mechanisms
    \end{itemize}
\end{enumerate}

As shown in the figures, both prompts were carefully structured to provide appropriate guidance while testing different capabilities. 
The zero-shot prompt emphasized comprehensive data context and clear coding constraints, while the agentic prompt focused on providing a structured workflow and tool interaction guidelines.

\begin{figure}[!t]
    \centering
    \begin{minipage}{0.95\columnwidth}
        \footnotesize
        \fbox{\begin{minipage}{0.95\columnwidth}
            \textbf{Zero-Shot Prompt Template:}\\
            Data Analysis Request\\
            Column names (attributes) of data to be analyzed:\\
            {[}Column names listed here{]}\\
            \\
            Sample row from the data (example):\\
            {[}Sample data row values listed here{]}\\
            \\
            Data description:\\
            {[}Detailed dataset description{]}\\
            \\
            {[}Optional instructional modules inserted here{]}\\
            \\
            Additional information: Don't use the sample row as input. Make sure to use only the corresponding column name(s) from the list provided above. Assume that you already have access to all the data stored in a variable named df. Don't use any variables other than df and those derived from df. Now do the following: Write Python code to {[}specific visualization task{]}. Provide a short description of the data, separated from the code.
        \end{minipage}}
    \end{minipage}
    \caption{The zero-shot system prompt template used in our experiments. 
    This template was populated with dataset-specific information and task requirements for each visualization request.}
    \label{fig:zero-shot-prompt}
\end{figure}

\begin{figure}[!t]
	\centering
	\begin{minipage}{0.95\columnwidth}
		\footnotesize
		\fbox{\begin{minipage}{0.95\columnwidth}
				\textbf{Algorithm 1: Agent Runner (pseudocode)}\\
				Input: user prompt, retriever, tools (list\_files, read\_file\_head, execute\_python\_code, read\_visualization\_image), max\_iters\\
				1: Retrieve top-1 dataset metadata and path\\
				2: For t = 1..max\_iters: plan → generate code → execute → read outputs/errors/image\\
				3: If plot valid and matches request: break; else revise based on feedback\\
				4: Save final code, plot, and logs
		\end{minipage}}
	\end{minipage}
	\caption{Pseudocode of the agentic loop.}
	\label{fig:agent-pseudocode}
\end{figure}

\begin{figure}[!t]
    \centering
    \begin{minipage}{0.95\columnwidth}
        \footnotesize
        \fbox{\begin{minipage}{0.95\columnwidth}
            \textbf{Agentic Prompt Template:}\\
            You are an expert data analysis and visualization assistant. Your goal is to help the user create a visualization based on their request.\\
            {[}Dataset context injected here{]}\\
            {[}Enabled modules injected here{]}\\
            
            You have access to a filesystem restricted to the './data/' directory. All outputs for this run will be saved within './output/{[}run\_id{]}/'.\\
            Key file paths for this run:\\
            - Log File: {[}log file path{]}\\
            - Executed Code: {[}code save path pattern{]}\\
            - Target Plot: {[}target output path{]}\\
            - Human Feedback: {[}feedback file path{]}\\
            
            Follow these steps:\\
            1. \textbf{Understand Request:} Clarify the user's goal.\\
            2. \textbf{Explore Data:} Use \texttt{list\_files} if needed.\\
            3. \textbf{Inspect Data:} Use \texttt{read\_file\_head}.\\
            4. \textbf{Plan Code:} Plan pandas/matplotlib/seaborn code.\\
            5. \textbf{Execute Code:} Use \texttt{execute\_python\_code}.\\
            6. \textbf{Analyze Results:} Check execution output.\\
            7. \textbf{Respond:} Explain steps, results, describe plot.\\
            
            Available Tools:\\
            - \texttt{list\_files}: Lists files in directory\\
            - \texttt{read\_file\_head}: Reads start of file\\
            - \texttt{execute\_python\_code}: Executes Python code\\
            - \texttt{read\_visualization\_image}: Reads the plot\\
            - \texttt{get\_human\_feedback}: Logs request for help
        \end{minipage}}
    \end{minipage}
    \caption{The agentic system prompt template used in our tool-equipped experiments. 
    This template guided the LLM through a structured workflow while providing access to specialized tools for data exploration, code execution, and visualization verification.}
    \label{fig:agentic-prompt}
\end{figure}
\vspace{-1em}

\subsection{Modular Prompt Components}

To systematically evaluate the impact of different instructional approaches, we implemented a modular prompt system with components that could be enabled or disabled:

\begin{enumerate}
    \item \textbf{Data Visualization Context}: Comprehensive principles for effective data visualization, covering core visualization principles, visual design rules, quality criteria, key considerations, and implementation guidelines.
    
    \item \textbf{Lessons Learned}: Domain-specific advice derived from previous visualization experiments, highlighting common errors in visual design, code implementation, and data representation to help the LLM avoid typical pitfalls.
    
    \item \textbf{Visualization Checklist}: A self-assessment framework prompting the LLM to evaluate its output against criteria such as clarity, data accuracy, and task alignment.
\end{enumerate}

These modules were implemented as optional text blocks that could be injected into the system prompts for both zero-shot and agentic experiments, allowing us to measure their impact on visualization quality and task success rates across different LLM providers and difficulty levels.

\subsection{Agentic System Architecture}

The implemented agentic visualization system consists of three main components:

\begin{enumerate}
    \item \textbf{Agent Runner}: Acts as the central orchestrator of the workflow. 
    The runner manages the conversation loop with the LLM, invokes tools based on the model's requests, establishes logging mechanisms, and maintains run-specific output directories for artifacts and performance tracking.
    
    \item \textbf{LLM Interface}: Facilitates communication with the underlying API. 
    This component handles sending conversation history, system prompts, and available tool specifications to the LLM, and processes its responses into structured formats (text responses, tool calls, and reasoning traces).
    
    \item \textbf{Tool Suite}: Provides the LLM with a set of specialized functions that enable interaction with the environment. 
    These tools allow for file system navigation, data inspection, code execution, visualization review, and human feedback solicitation when necessary.
\end{enumerate}

File access is restricted to predefined directories (\texttt{data/} for input, \texttt{output/} for generated files) to ensure experimental control and reproducibility.
\subsection{Agent Workflow}

\begin{figure}[!t]
    \centering
    \includegraphics[width=0.85\columnwidth]{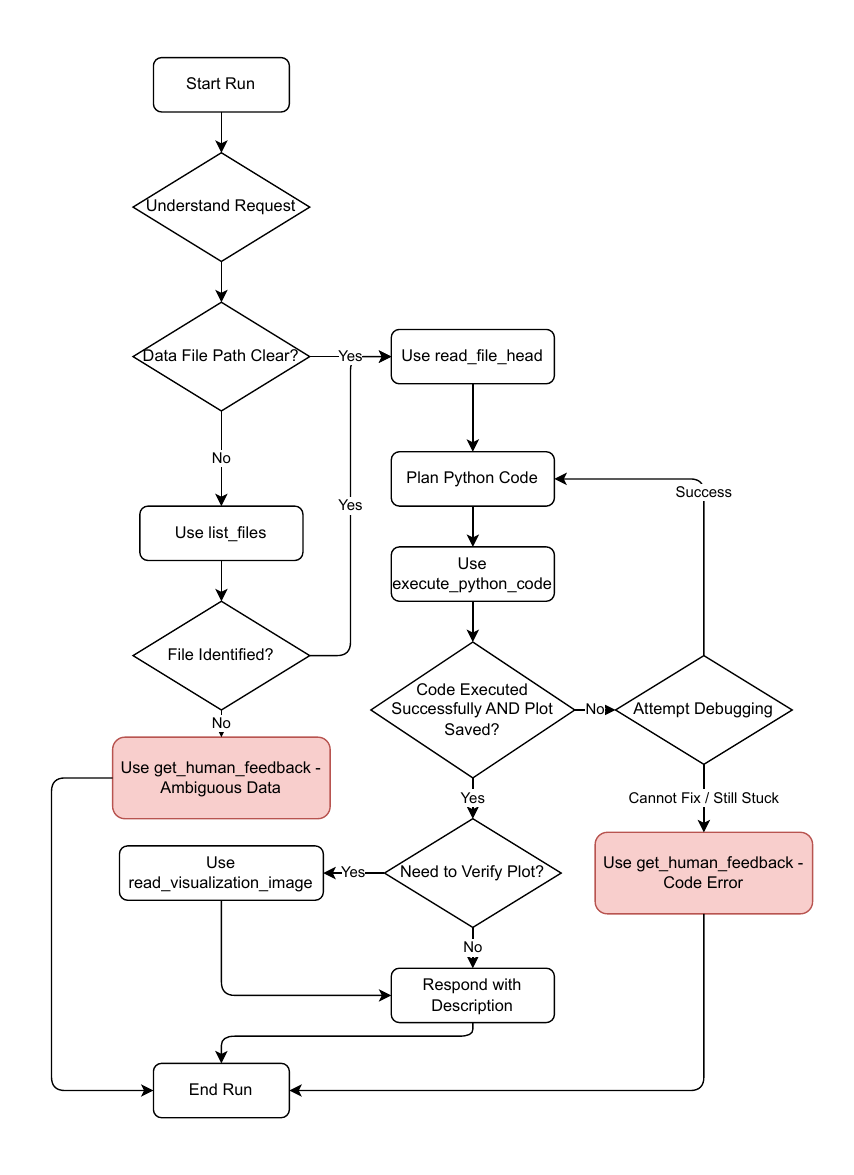}
    \caption{Typical decision flow of the agentic visualization system.}
    \label{fig:agent-workflow}
\end{figure}

The agentic system's workflow, depicted in Figure \ref{fig:agent-workflow}, follows a structured sequence that enables iterative visualization development:

\begin{enumerate}
    \item \textbf{Request Understanding}: Comprehend and clarify the user's visualization goal
    
    \item \textbf{Data Exploration}: Navigate the filesystem to locate relevant datasets
    
    \item \textbf{Data Inspection}: Examine data structure and content to inform visualization planning
    
    \item \textbf{Code Planning}: Develop a strategy for implementing the visualization using Python libraries
    
    \item \textbf{Code Execution}: Generate and execute Python code with explicit file paths and output locations
    
    \item \textbf{Result Analysis}: Evaluate execution outcomes, debug errors, and verify visualization output
    
    \item \textbf{Response Generation}: Explain process, describe results, or request human intervention when necessary
\end{enumerate}

\subsection{LLM-Agent Capabilities}

The LLM-based agent is equipped with five primary capabilities through its tool interface:

\begin{enumerate}
    \item \textbf{Data Discovery}: Using the \texttt{list\_files} tool, the agent can navigate through the available datasets in the data directory to identify relevant files for a given visualization request.
    
    \item \textbf{Data Inspection}: Through the \texttt{read\_file\_head} tool, the agent can examine the first several lines or rows of a dataset to understand its structure, variables, and content before attempting visualization.
    
    \item \textbf{Code Generation and Execution}: The \texttt{execute\_python\_} \texttt{code} tool enables the agent to generate and run Python code using standard data science libraries (pandas, matplotlib, seaborn, numpy) to produce visualizations. 
    The system automatically preserves each code iteration for analysis.
    
    \item \textbf{Visualization Verification}: Using  \texttt{read\_visualization\_} \texttt{image}, the agent can inspect the generated visualization, verify its correctness, and potentially refine it based on the observed output.
    
    \item \textbf{Assistance Seeking}: When faced with unresolvable errors or ambiguities, the agent can use the \texttt{get\_human\_feedback} tool to document its reasoning process and request human intervention. 
    This tool was never invoked in practice.
\end{enumerate}

\subsection{Design Rationale}

We briefly justify key design decisions. After testing smaller sentence transformers, we decided to use \texttt{all-mpnet-base-v2} as it handled domain terms (e.g., labour) more robustly than lighter models, and outperformed sparse BM25 on semantic matches under short prompts. We study both a zero-shot setting (one-pass code) and an agentic setting to quantify the incremental value of iterative execution and self-correction. The tool suite is deliberately minimal (file listing, head inspection, code execution, image read) to mirror a realistic sandbox and reduce degrees of freedom.

\section{Evaluation Methodology\label{sec:evaluation}}

\subsection{Experimental Protocol}

Our experimental evaluation followed a structured protocol to systematically assess the agent's visualization capabilities:

\begin{enumerate}
    \item \textbf{Task Selection}: We compiled a diverse set of visualization tasks based on official statistics datasets from Statistics Netherlands, varying in complexity from simple time series plots to multi-variable comparative visualizations. 
    Tasks were categorized into three difficulty levels:
    \begin{itemize}
        \item \textbf{Easy}: Simple single-variable visualizations (e.g., ``Plot the volume of cheese production in the Netherlands'')
        \item \textbf{Medium}: Tasks requiring data filtering or multiple variables (e.g., ``Plot the monthly volume of raw cow's milk delivered by dairy farmers between 2010-2015'')
        \item \textbf{Hard}: Complex tasks with specific analytical requirements (e.g., ``Plot the seasonally adjusted daily turnover for domestic and foreign markets for sector '16 Manufacture of wood products' between 2020-2021'')
    \end{itemize}
    
    \item \textbf{Dataset Preparation}: We utilized seven real-world official statistics datasets covering diverse domains:
    \begin{itemize}
        \item Milk supply and dairy production by factories
        \item Caribbean Netherlands births, fertility, and age of mother
        \item Consumer price indices
        \item Industry production and sales statistics
        \item Producer price indices
        \item Municipal accounts balance sheets
        \item Population demographics by sex, age, generation, and migration background
    \end{itemize}
    These datasets were selected to ensure broad coverage of both categorical and numerical data types, enabling the generation of a wide variety of visualization styles such as bar charts, line graphs, pie charts, heat maps, etc. The selection process also considered the structural complexity and ambiguity of attributes to evaluate how well LLMs manage real-world challenges. In addition to technical variability, usage metrics provided by Statistics Netherlands, such as visitor counts, session frequency, and returning user ratios, were analyzed to prioritize datasets frequently accessed by the public. This ensured that the evaluation and prompt refinement processes remained closely aligned with actual user behavior and practical relevance.
    
    \item \textbf{Experimental Approaches}: We implemented two distinct approaches to evaluate and compare LLM capabilities:
    \begin{itemize}
        \item \textbf{Zero-Shot Generation}: Tasks were presented to LLMs with dataset context and visualization instructions, requiring the model to generate a complete Python visualization solution in a single response without iteration or feedback.
        
        \item \textbf{Agentic Workflow}: The same tasks were presented to LLMs equipped with tools for data exploration, code execution, visualization verification, and error recovery, enabling an iterative development process similar to human data scientists.
    \end{itemize}
    
    \item \textbf{Provider Comparison}: We conducted experiments across multiple LLM providers (including Anthropic, DeepSeek, Google, OpenAI, and others) to assess whether capabilities varied by model architecture and training approach. See Table~\ref{tab:llm_results_all} for the list of included models with version numbers.
    
\end{enumerate}

\subsection{Scoring framework}
Model outputs were inspected along three categories: \begin{itemize} \item \textbf{Visual quality} – purely graphical concerns, including axis placement, tick formatting, label legibility, colour or marker choices, and general adherence to design best-practice. \item \textbf{Data correctness} – verification that the figure reflects the prompt: correct columns must be chosen, filters must be in line with any stated conditions, and aggregations (e.g., sums or year-on-year values) must be applied exactly once. \item \textbf{Code reliability} – assessment of the Python script itself: it should run without errors, operate on the provided \texttt{df} rather than hard-coded literals, and avoid redundant computations. \end{itemize}
To keep grading consistent across 200 prompt–model pairs we converted each requirement into a Yes/No item, yielding a short binary checklist for every category.
Binary scoring was specifically selected as it limits subjectivity, speeds up evaluation, and lets us compute average hit-rates per model with simple aggregation.

Examples include ``Are the axis labels clear and readable?'' (visual), ``Does the filtering process correctly reflect the user’s intent?'' (data), and ``Does the code execute without syntax errors and successfully generate a graph?'' (code).
An answer of \emph{Yes} scores 1, \emph{No} scores 0. Then, category totals are normalised to scores out of 10 to obtain visual, data, and code sub-scores, which are then combined into an overall quality score.
A short version of the complete checklist used in our study is shown in Table \ref{tab:case_study}. For the full version, please refer to the appendix at \url{https://github.com/aida-ugent/llm_visualization_results}.

\section{Results\label{sec:results}}

Before presenting the overall scores in Section~\ref{sec:quantitativeresults}, we make these results more tangible by first presenting a few example outcomes in Section~\ref{sec:casestudyresults} and then an overview of various problems that occurred in the outputs in Section~\ref{sec:qualitativeresults}.

\subsection{Case Study\label{sec:casestudyresults}}

\begin{figure}[!t]
    \centering
    \begin{tabular}{cc}
        \includegraphics[width=0.45\linewidth]{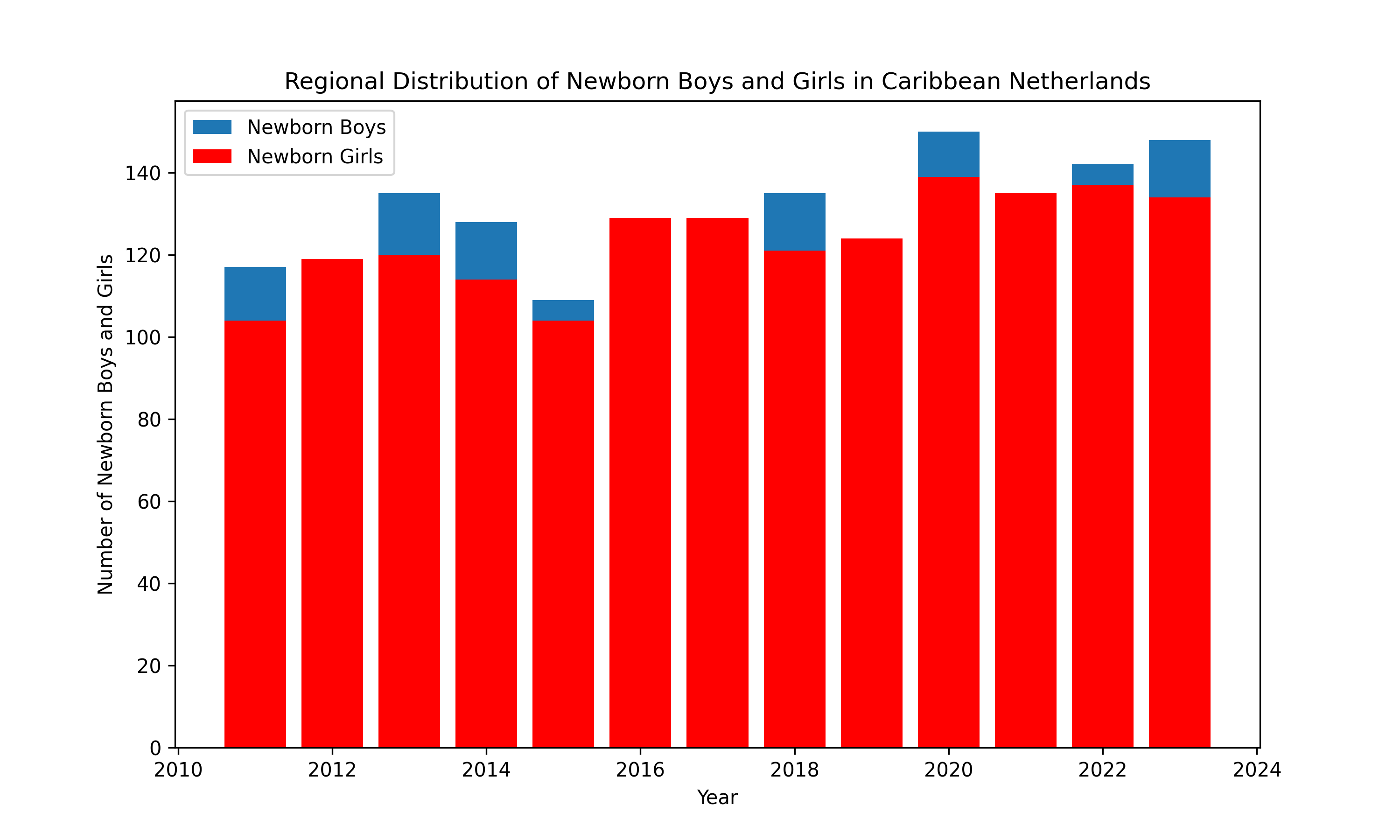} & 
        \includegraphics[width=0.45\linewidth]{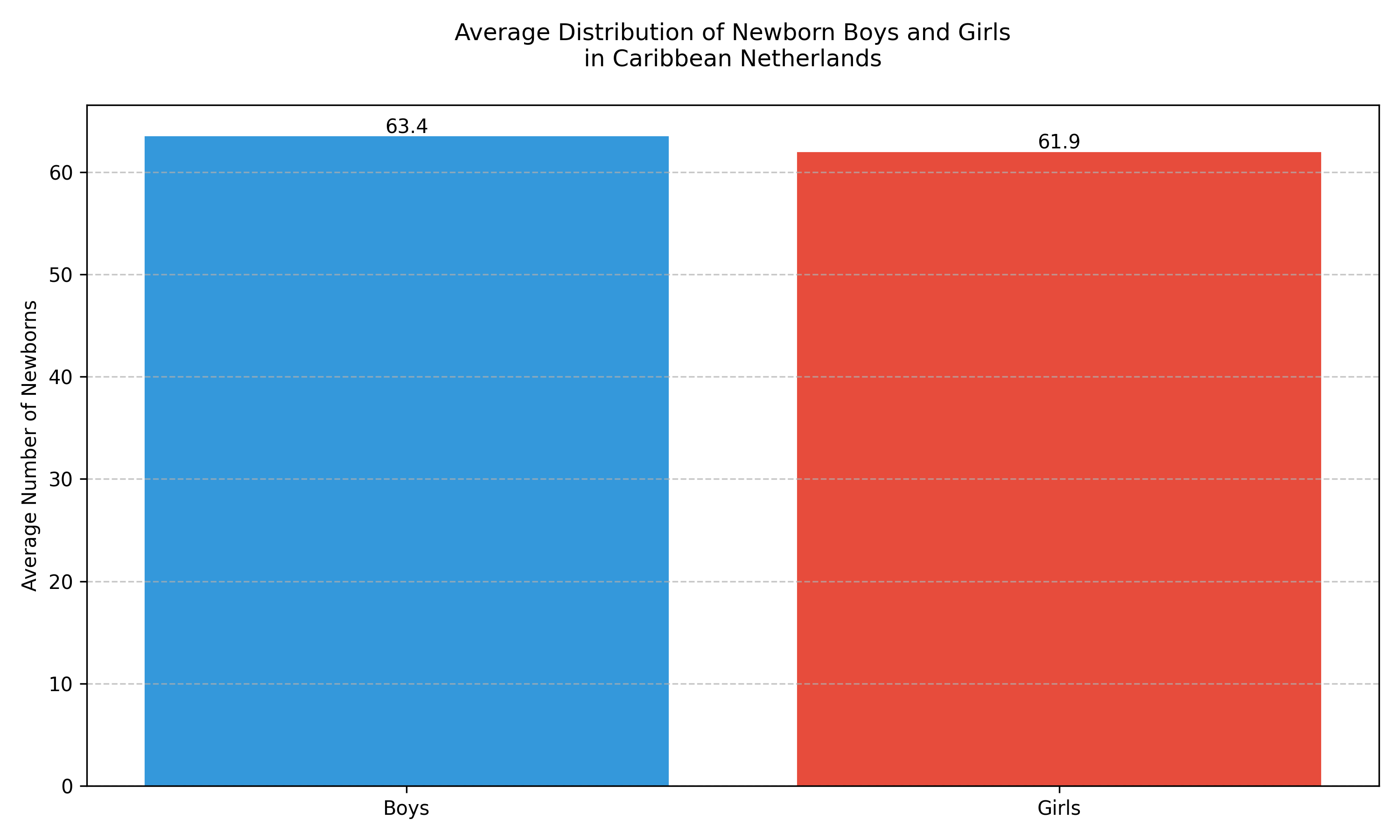} \\
        \small (a) Llama3.1 & \small (b) Claude 3.5 \\
        \includegraphics[width=0.45\linewidth]{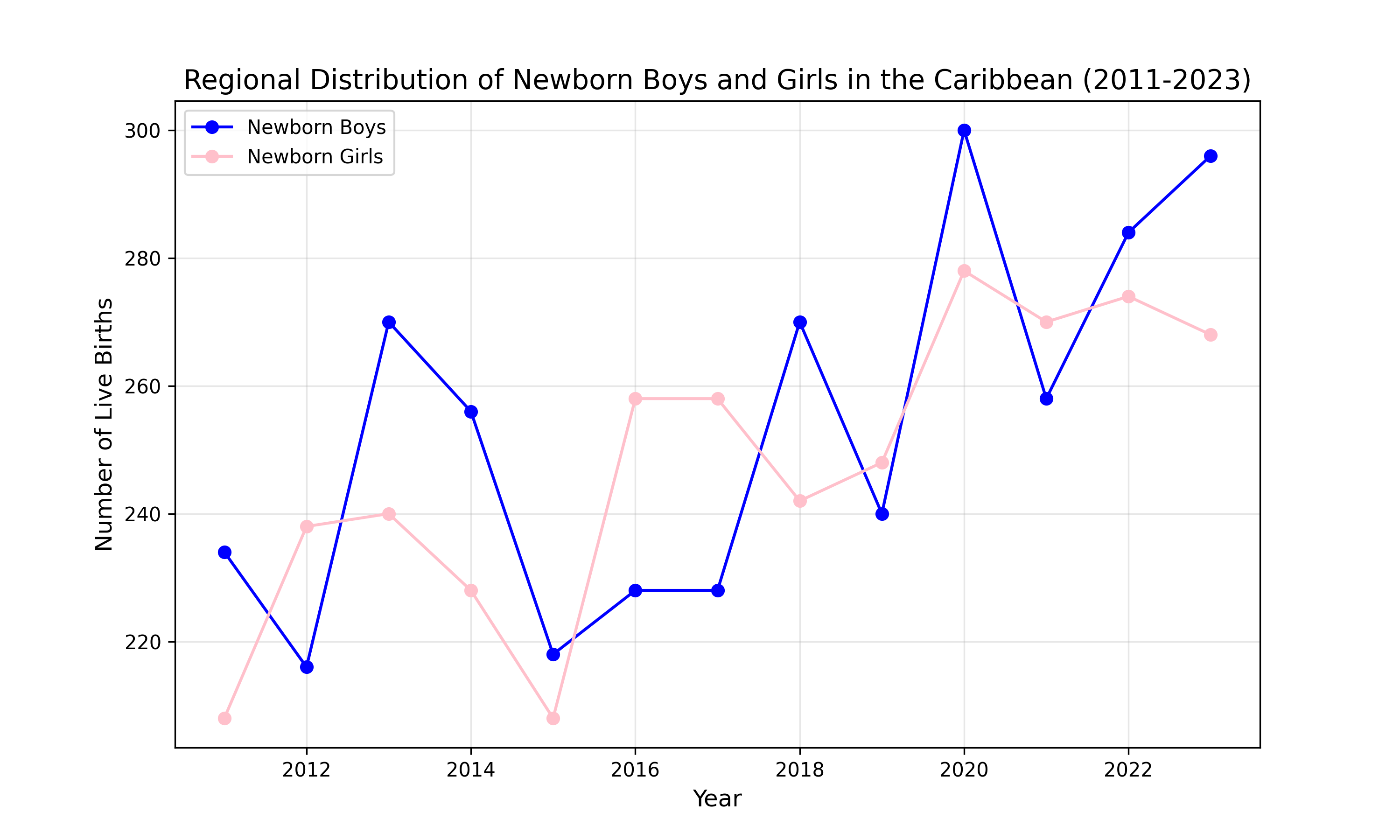} & 
        \includegraphics[width=0.45\linewidth]{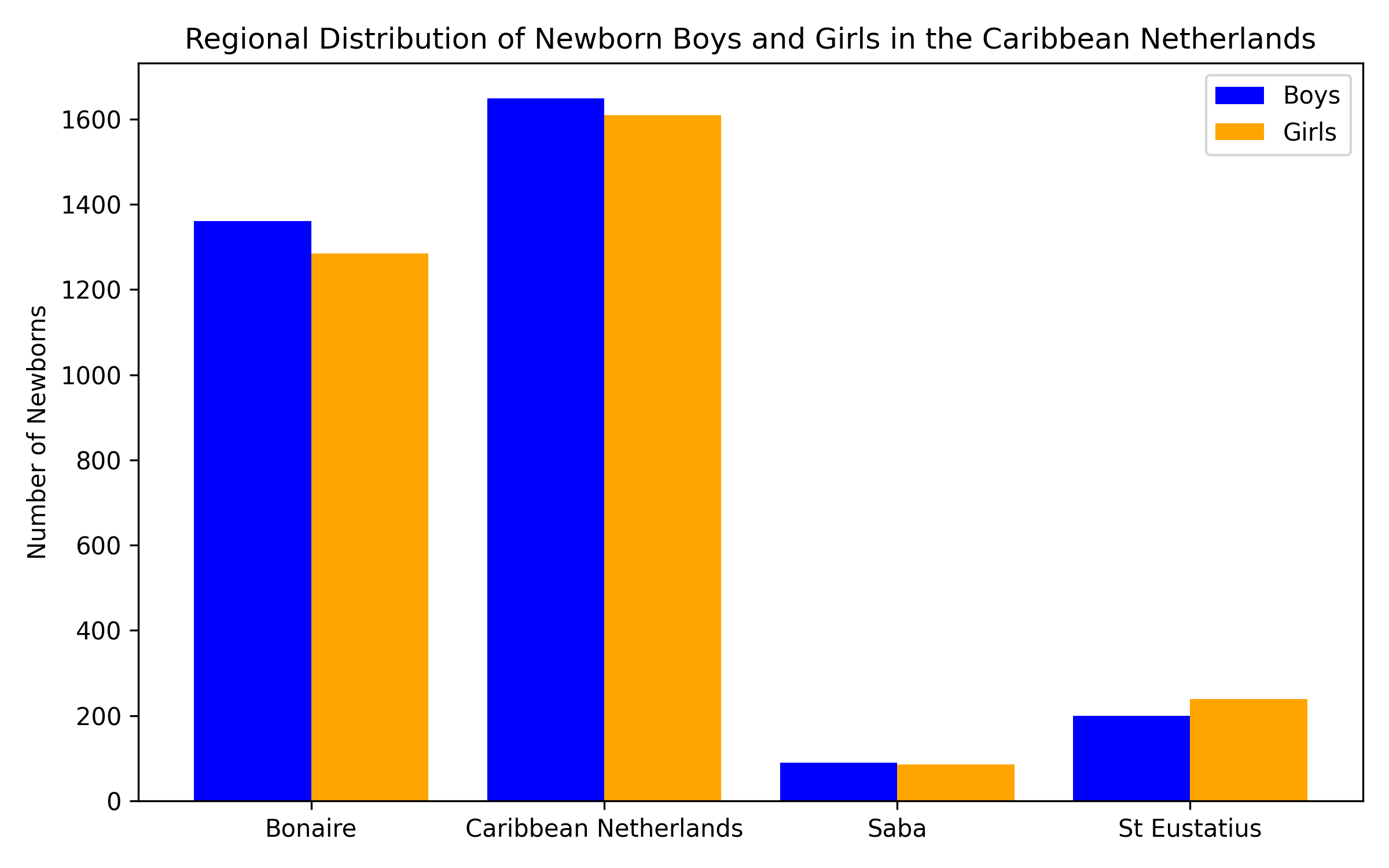} \\
        \small (c) GPT-4 & \small (d) o1-High \\
    \end{tabular}
    \caption{Visualizations generated by different LLMs for the prompt: \textit{``Plot the regional distribution of newborn boys and girls in Caribbean.''}}
    \label{fig:caribbean_llm_comparison}
\end{figure}

Figure~\ref{fig:caribbean_llm_comparison} shows the results from four base (non-agentic) LLMs to the same prompt, illustrating how different LLMs interpret the same task. This prompt required models to compare the number of newborn boys and girls across different regions in the Caribbean Netherlands. We briefly discuss in which ways their outputs fail or succeed, and how visual, code, and data errors are reflected in the scoring framework.

\textbf{Llama3.1} generated a bar chart with two traces overlayed over each other. The bar representing boys is not visible when smaller in value than girls. More critically, the model misinterpreted the grouping intent of the prompt by aggregating over years rather than regions, likely defaulting to temporal logic when encountering time-related fields. This resulted in deduction of scores due to incorrect filtering and column selection.

\textbf{Claude 3.5} executed the code correctly and avoided syntax issues, but failed to aggregate the data by region entirely. Instead, it produced a single averaged result for the entire Caribbean Netherlands, overlooking geographic distinctions explicitly requested in the prompt.

\textbf{GPT-4} introduced two major issues: it aggregated by year instead of region and used a line chart for categorical data, which was not an appropriate encoding. These choices resulted in visual misalignment and semantic mismatch.

\textbf{o1-High}, by contrast, successfully aggregated the data by region and plotted grouped bars for boys and girls side-by-side, ensuring both visibility and comparability, achieving the highest score in all three categories. The evaluation table detailing category-wise scoring for this prompt is included in Table \ref{tab:case_study}.

\begin{table}[!t]
\caption{Case study evaluation results.}.
\label{tab:case_study}
\begin{tabular}{@{}lcccc@{}}
\toprule
\textbf{Criteria} & \textbf{Llama} & \textbf{Claude} & \textbf{GPT-4} & \textbf{o1} \\
\midrule
\multicolumn{5}{@{}l}{\textbf{Visual Quality Scores}} \\
\hline
X-axis correct        & 1 & 1 & 1 & 1 \\
Y-axis correct        & 1 & 1 & 1 & 1 \\
Axis labels clear     & 1 & 1 & 1 & 1 \\
Color used well       & 1 & 1 & 1 & 1 \\
Legend accurate       & 1 & 1 & 1 & 1 \\
Good scaling          & 1 & 0 & 0 & 1 \\
Marks correct         & 0 & 1 & 1 & 1 \\
Readable layout       & 1 & 1 & 1 & 1 \\
\textbf{Total}   & \textbf{7} & \textbf{7} & \textbf{7} & \textbf{8} \\
\midrule
\multicolumn{5}{@{}l}{\textbf{Code Quality Scores}} \\
\hline
Correct imports       & 1 & 1 & 1 & 1 \\
Code runs             & 1 & 1 & 1 & 1 \\
Correct columns       & 0 & 1 & 1 & 1 \\
Filters correctly     & 0 & 1 & 1 & 1 \\
No hardcoding         & 1 & 1 & 1 & 1 \\
Prompt fully handled  & 0 & 1 & 0 & 1 \\
No redundancy         & 1 & 1 & 1 & 1 \\
\textbf{Total}   & \textbf{4} & \textbf{7} & \textbf{6} & \textbf{7} \\
\midrule
\multicolumn{5}{@{}l}{\textbf{Data Accuracy Scores}} \\
\hline
Correct chart type    & 1 & 1 & 0 & 1 \\
Column selection      & 0 & 1 & 1 & 1 \\
Correct filtering     & 1 & 0 & 0 & 1 \\
Correct aggregation   & 1 & 0 & 0 & 1 \\
Subset accurate       & 0 & 1 & 1 & 1 \\
Handles nulls         & 0 & 1 & 1 & 1 \\
Prompt fully covered  & 0 & 1 & 0 & 1 \\
\textbf{Total}   & \textbf{3} & \textbf{5} & \textbf{3} & \textbf{7} \\
\bottomrule
\end{tabular}
\end{table}

\subsection{Qualitative Results\label{sec:qualitativeresults}}

Beyond the specific issues highlighted in the case study, a broader analysis of outputs, particularly from zero-shot generation attempts without iterative feedback, reveals consistent challenges across models and prompt types. 
These recurring problems, categorized into visual design, code implementation, and data logic, underscore the limitations of unassisted LLM-based visualization and motivated the development of more sophisticated approaches.
\subsubsection{Visual Representation Issues}

A frequently observed issue was the misuse of categorical x-axes, especially when plotting the \texttt{Periods} column, which is stored as strings. This led to disordered axes and uneven spacing since plotting libraries like \texttt{matplotlib} do not automatically interpret strings as temporal data. Another common flaw was failing to anchor the y-axis at zero, resulting in graphs that exaggerated minor fluctuations.

LLMs also tended to add redundant elements like both lines and markers in the same plot, which cluttered the visual and reduced clarity. Furthermore, strong gridlines and dense background structures were often included by default, diminishing the visual focus on the data. These patterns suggest that while models can produce syntactically valid plots, they lack intuition for effective visual communication.

\begin{figure}[!t]
\centering
\includegraphics[width=0.75\linewidth]{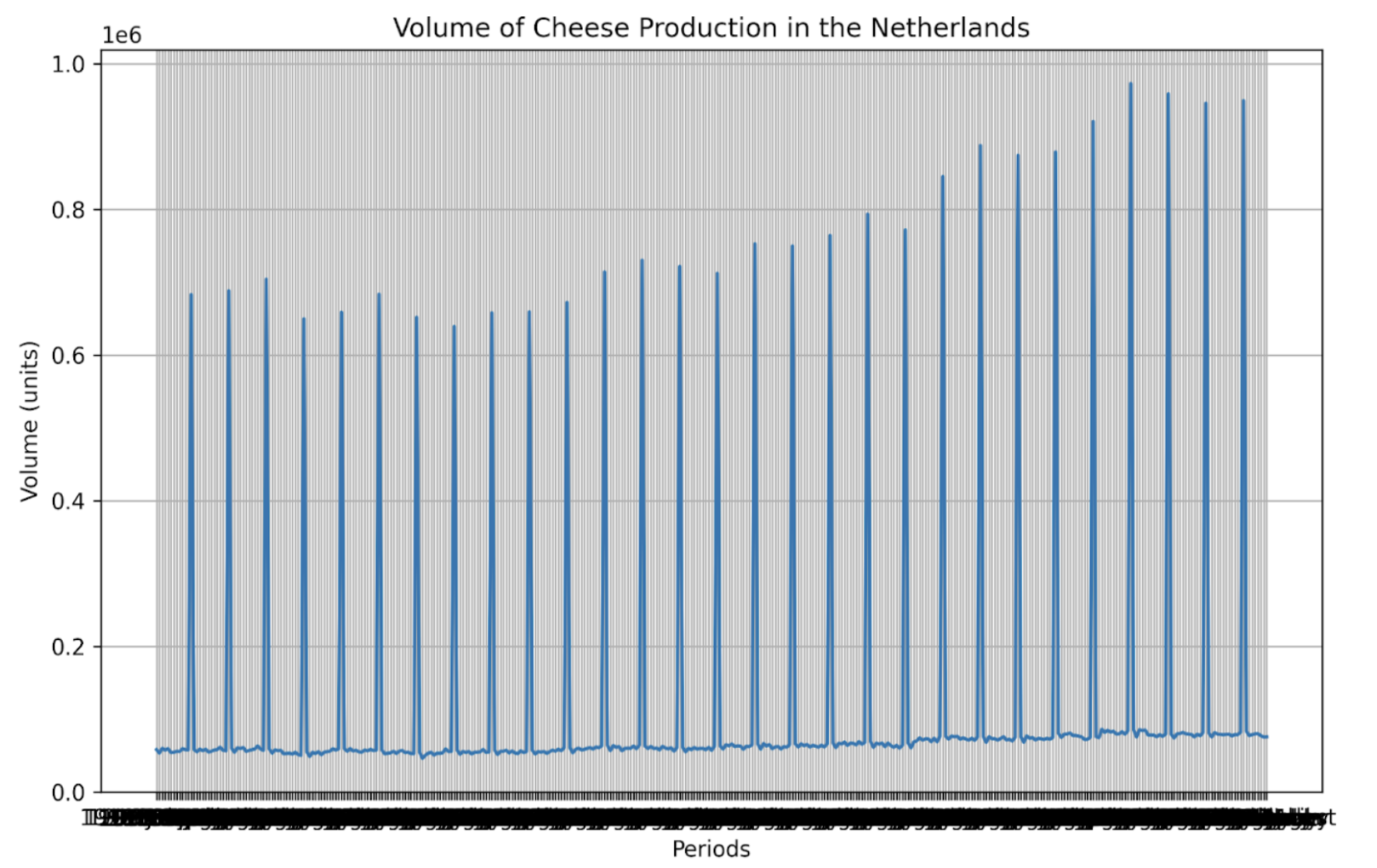}
\caption{Visualization with poor axis sorting and excessive gridlines—typical issues seen across several models.}
\label{fig:visual_problem_graph}
\end{figure}

\subsubsection{Code Generation Issues}

Several LLMs incorrectly attempted to convert non-standard date strings into datetime objects, causing errors during execution. A broader issue was the tendency to guess column names or refer to nonexistent fields due to the lack of direct data inspection. Hardcoding values—especially prevalent in models like Gemini 2.0 Flash Thinking—was another severe flaw, undermining prompt flexibility and generalization.

Filtering logic was often mishandled, either by applying incorrect filters or omitting them altogether. These shortcomings stem largely from the model’s inability to validate its assumptions against real data, leading to brittle code that may run but not meet the task requirements.

\subsubsection{Data Representation Issues}

Some models selected inappropriate chart types—for instance, using line graphs where bar charts were more suitable. Missing aggregations also led to visual noise, especially when plotting multi-period or multi-region data without combining values logically.

In other cases, null values and outliers were left unprocessed, resulting in jagged or misleading trends. Additionally, axis labels sometimes contained raw codes or identifiers from the dataset, making the plots less accessible for human interpretation.

\begin{figure}[!t]
\centering
\includegraphics[width=0.75\linewidth]{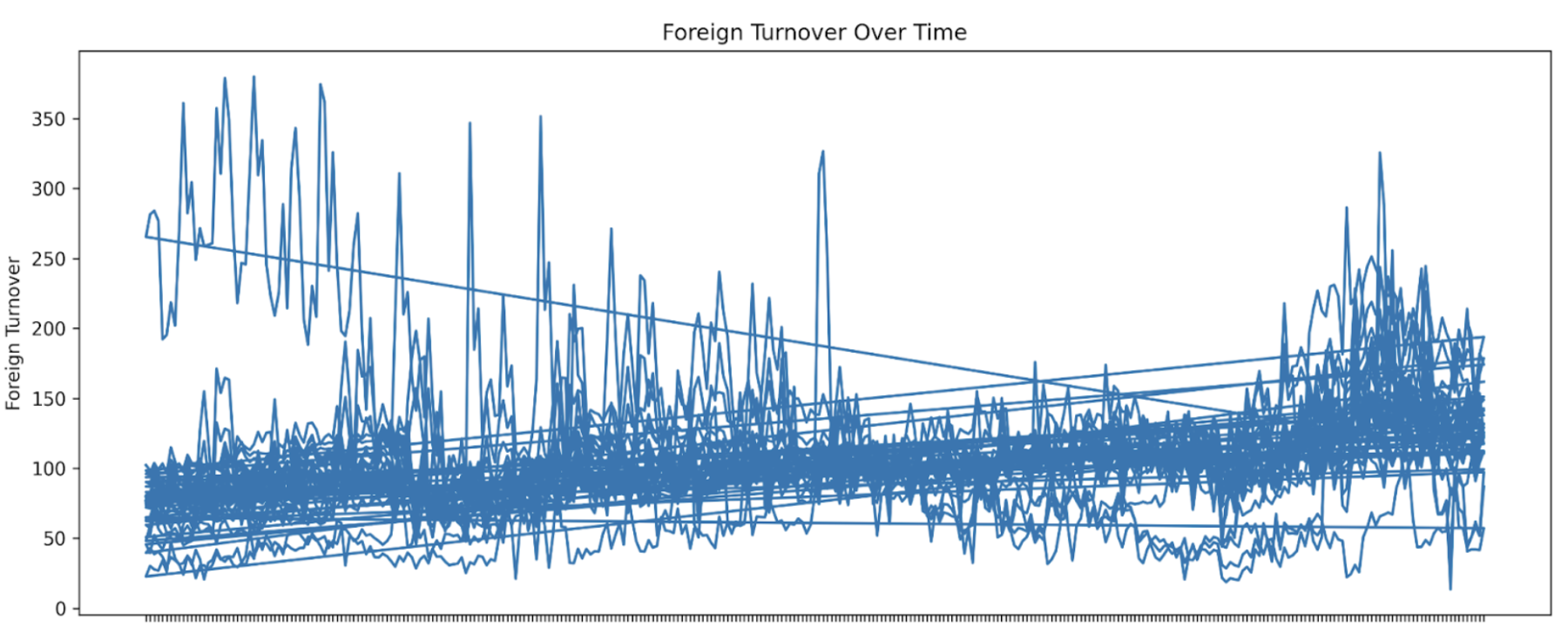}
\caption{A case where failure to aggregate on the \texttt{Periods} column results in overlapping lines.}
\label{fig:data_problem_graph}
\end{figure}

\subsubsection{Issues Beyond LLM Capabilities}

Not all problems stem from the language models themselves. Some are rooted in the default behavior of tools like \texttt{matplotlib}, which often produce dense grid structures or fail to sort string-based axes. Others arise from the structure of the dataset, such as long column names with embedded indices or special characters that are difficult to reference programmatically.

These issues underscore the importance of pre-processing steps, such as renaming columns, cleaning missing values, or formatting dates before passing data to the model. More intelligent defaults in plotting libraries or clearer prompt guidance could potentially mitigate these problems.

These issues highlight the need for mechanisms that allow LLMs to refine their outputs, validate assumptions, and adhere more closely to visualization best practices. 

\subsection{Quantitative Results\label{sec:quantitativeresults}}

To systematically assess LLM performance and the impact of different prompting strategies and system designs, we conducted a quantitative evaluation. We evaluated 11 model configurations, including 8 base LLMs and 3 improved variants, across 25 prompts. The prompts are categorized into 7 Easy (involving single-variable plots), 11 Medium (including multiple variables or filtering steps), and 7 Hard (requiring more complex logic and multi-dimensional comparisons) difficulty levels. The evaluation criteria involved 22 binary questions, which are grouped into three main dimensions: Visual (8 questions), Code (7 questions), and Data (7 questions). Each question was scored with a 1 (Yes) or 0 (No), and category scores were normalised to a 10-point scale, as discussed earlier.

Table~\ref{tab:llm_results_all} presents the aggregated results for all models and configurations. 
\begin{table}[!t]
\centering
\caption{Normalised evaluation scores (out of 10), averaged over the 25 prompts, for each model and configuration.}
\label{tab:llm_results_all}
\begin{tabular}{|l|c|c|c|}
\hline
\textbf{Model/Configuration} & \textbf{Visual} & \textbf{Code} & \textbf{Data} \\
\hline
Llama3.1 (9B) & \textit{5.95} & 8.00 & \textit{4.86} \\
Gemma 2 (9B) & \textit{5.55} & 8.23 & \textit{4.91} \\
Qwen 2.5 (7B) & 5.85 & 8.63 & 5.71 \\
Claude 3.5 & 6.90 & 8.23 & 6.69 \\
Deepseek-Chat & 7.00 & 8.57 & 6.74 \\
Gemini 2.0 Flash Thinking & 7.05 & 7.77 & 6.00 \\
GPT-4o & 6.20 & 7.94 & 6.00 \\
o1-High & 7.50 & \textbf{8.97} & \textbf{7.37} \\
\textit{o1-High + Context} & \textbf{7.80} & 8.91 & 7.03 \\
\hline
\textit{Claude 3.7 + Feedback Loop} & 8.90 & \textbf{9.83} & \textbf{9.43} \\
\textit{Claude 3.7 + Feedback + Context} & \textbf{9.05} &  9.71 & \textbf{9.43} \\
\hline
\end{tabular}
\end{table}
Across the base models, code generation was the strongest category. All models except Gemini 2.0 Flash Thinking scored above 8.0. Notably, Qwen 2.5, with 7B parameters, achieved an impressive 8.63, outperforming much larger models like Claude 3.5 (8.23) and GPT-4 (7.94). The best-performing base model in this category was o1-High with 8.97. This suggests that model size alone does not determine code reliability; smaller models can compete effectively when pretrained on high-quality code data.

Data handling proved to be the most challenging dimension for all of the models. The two smallest models, Llama3.1 (9B) and Gemma 2 (9B), had the weakest performance here, scoring just 4.86 and 4.91, respectively. These errors often arose from missing or incorrect filters, grouping mistakes, or improper column selection. In contrast, o1-High achieved the top base score of 7.37, followed by Deepseek-Chat (6.74) and Claude 3.5 (6.69), showing a stronger ability to manipulate structured datasets.

For most models we observed a performance drop as prompt difficulty increased, especially in the data reasoning aspect. Only Claude 3.5 and Deepseek-Chat deviated from this trend, with higher scores on more complex prompts, indicating better generalization under increasing logical complexity. More details on the breakdown of scores based on difficulty can be found in the appendix at \url{https://github.com/aida-ugent/llm_visualization_results}. 

Visual output quality varied most widely across models. Gemini (7.05), Deepseek-Chat (7.00), and o1-High (7.50) produced mostly consistent and well-formatted charts. In contrast, Llama3.1 and Gemma 2 scored poorly in this category (5.95 and 5.55), often due to overlapping labels, incorrect axes, or inappropriate chart types, such as using line charts for categorical comparisons, as shown in the case study.

Adding structured context to the prompt, covering visualization principles, common pitfalls, and a checklist led to an improvement in o1-High's visual score, increasing from 7.50 to 7.80. However, its code and data scores slightly declined to 8.91 (from 8.97) and 7.03 (from 7.37). This suggests that while visual clarity improved, the added instructions may have introduced complexity that confused the model in areas requiring precise logic or filtering.
\begin{figure}[!t]
    \centering
    \includegraphics[width=0.9\columnwidth]{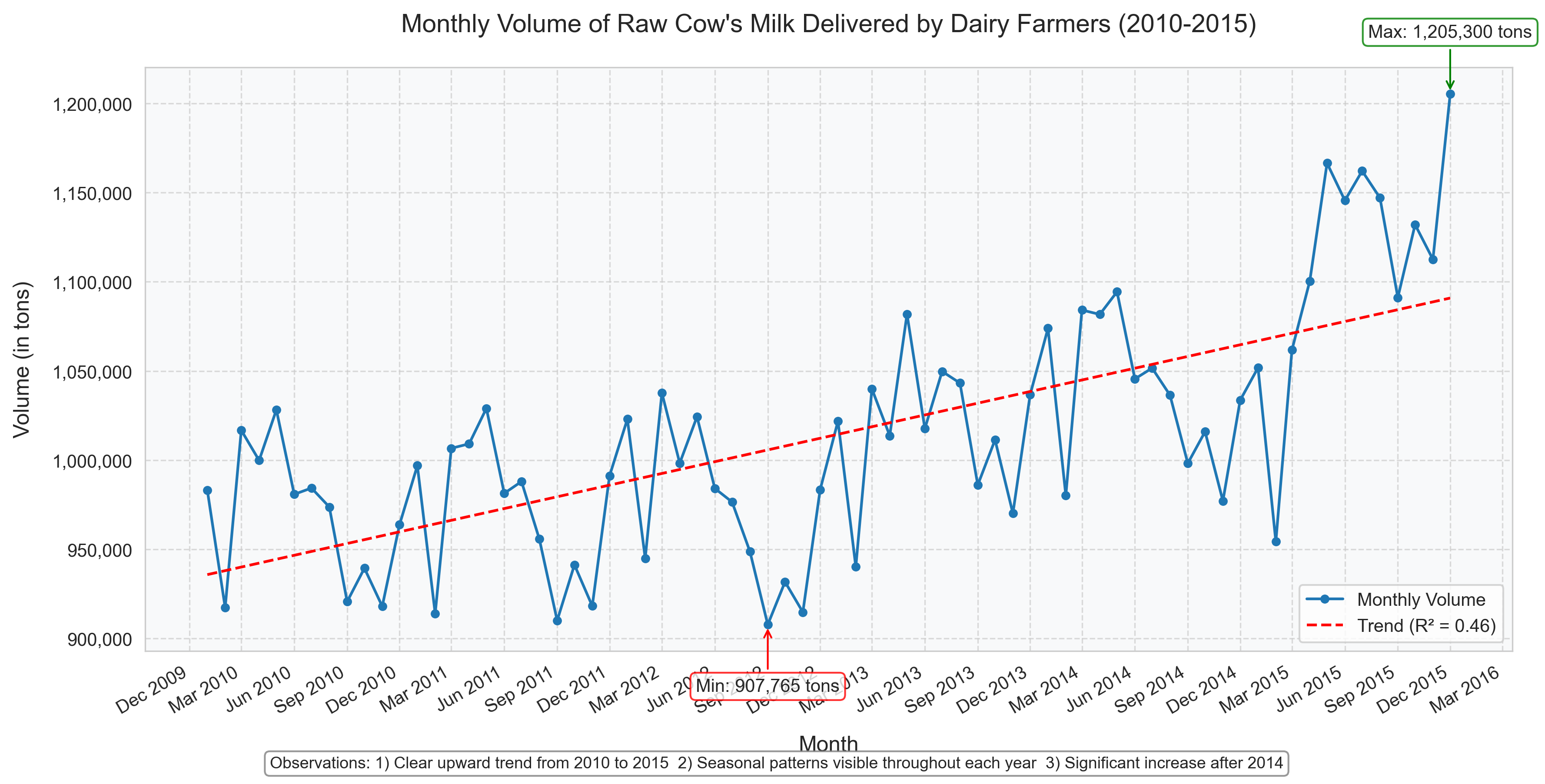}
    \caption{\textit{``Plot the monthly volume of raw cow's milk delivered by dairy farmers between 2010–2015''}. Generated by Claude 3.7 with feedback loop and contextual prompt guidance. The line chart accurately captures seasonal fluctuations. It also includes a red trendline, annotated min and max values, and an interpretive legend summarizing observed patterns.}
    \label{fig:claude_milk}
\end{figure}

Enabling Claude 3.7 to revise its own outputs through a 25-step feedback loop resulted in dramatic improvements across all categories. Its visual, code, and data scores jumped to 8.90, 9.83, and 9.43, respectively—among the highest in the evaluation. The model benefitted from the ability to inspect its own plots, correct execution errors, and adapt its strategy over time.

\begin{figure}[!t]
    \centering
    \includegraphics[width=0.9\columnwidth]{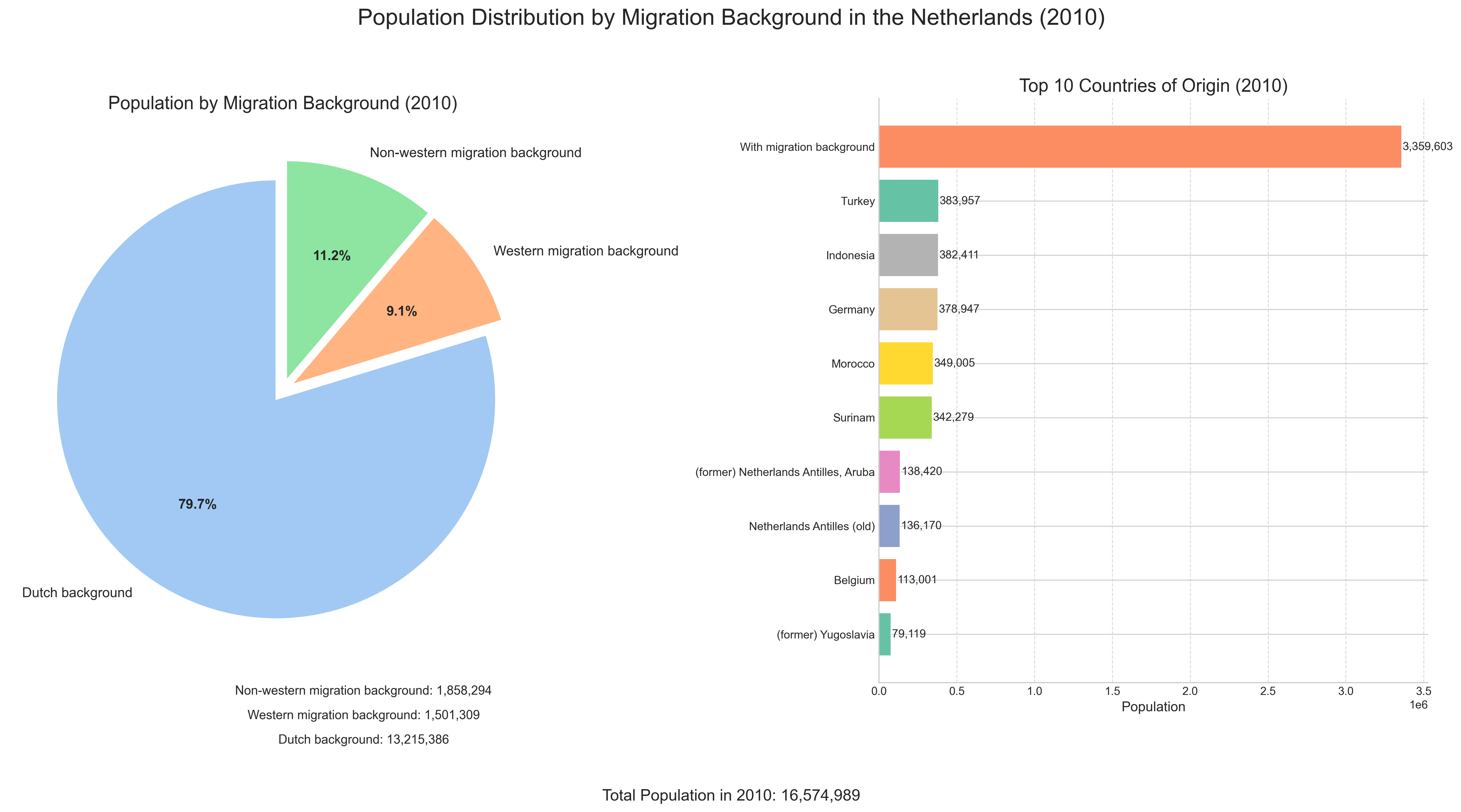}
    \caption{\textit{``Plot population distribution by migration background in the year 2010.''} This dual-plot visualization includes a pie chart and horizontal bar chart for complementary perspectives. It clearly distinguishes the main population groups and ranks the top 10 origin countries by population with precise values, labels and consistent color coding.}
    \label{fig:claude_population}
\end{figure}

Combining both enhancements (prompt context and self-feedback) pushed Claude’s visual score even further to 9.05, while maintaining high performance in code (9.71) and data (9.43). This final configuration produced the best overall scores in the study, demonstrating that iterative refinement and additional context in prompts could be used to reinforce the performance of the models. Two of such visualizations are provided in Figures \ref{fig:claude_milk} and \ref{fig:claude_population}.

In summary, these findings highlight several key observations:
\begin{itemize}
    \item All base models perform reasonably well in code generation, but smaller models (7B–9B) struggle with data logic and visual design.
    \item o1-High consistently outperforms other base models, achieving the highest scores in visual, code, and data dimensions.
    \item Adding prompt-level context improves visual presentation, though it may slightly reduce code or data accuracy.
    \item Iterative feedback mechanisms enable substantial gains, especially for complex prompts requiring layered reasoning.
    \item When self-feedback is combined with context guidance, models like Claude 3.7 reach near-perfect scores, indicating the full potential of LLM agents in visualization tasks.
\end{itemize}

Full analysis table for each model (per question, per prompt) is included in the appendix at \url{https://github.com/aida-ugent/llm_visualization_results}.

\section{Conclusions \& Discussion\label{sec:discussion}}

This study evaluated the capability of large language models to generate data visualizations from official statistics data. 
Through systematic experiments with eight LLMs across 25 tasks, we found that while base models, under one-shot conditions, achieve adequate code generation (mean score: 8.3/10), they exhibit substantial deficiencies in data manipulation (5.9/10) and visual design (6.5/10).  
The implementation of an agentic system with iterative self-correction yielded dramatic improvements, with Claude 3.7 achieving scores exceeding 9.0/10 across all three dimensions.

While the visualizations generated by our agentic system nearly saturate our benchmark, they suffer from an increase in complexity of the produced charts that our evaluation did not explicitly take into account.
Care needs to be taken that these efforts by LLMs to iterate on their own results do not push their results outside of the grasp of the intended audience.

Beyond these technical findings, this work makes several broader contributions. 
We introduce a reusable evaluation framework for systematically assessing LLM performance across visualization tasks of varying complexity. 
By separating evaluation into three dimensions (code, data, visual), our approach provides a replicable template for future benchmarking efforts in the field. 
The generic nature of the agentic approach developed in this study suggests broader applicability beyond the public data of Statistics Netherlands. 
The modular prompt engineering, evaluation framework, and iterative self-correction mechanisms are domain-agnostic and could potentially be extended to other statistical agencies, corporate dashboards, scientific publications, and educational contexts. 

The findings also offer actionable insights into how LLMs reason through structured tasks, highlighting where modifications in prompt engineering or model design could yield improvements.
More broadly, this research contributes to the growing field of natural language interfaces (NLI) to data visualization---also referred to as text-to-vis---by demonstrating how advanced language models can make complex data more accessible to a broader audience.

It could be studied in further detail whether the generated visualizations are a good fit for a non-expert audience. Conducting comparative user studies between LLM-generated visualizations and those produced by Statistics Netherlands experts would provide insights into practical comprehensibility for non-expert users. 
Such studies could establish whether the technical sophistication we observed enhances or hinders the democratization of official statistics access.

\begin{acks}
The research leading to these results has received funding from the Special Research Fund (BOF) of Ghent University (BOF20/IBF/117), from the Flemish Government under the ``Onderzoeksprogramma Artificiële Intelligentie (AI) Vlaanderen'' programme, from the FWO (project no. G0F9816N, 3G042220, G073924N). Funded/Co-funded by the European Union (ERC, VIGILIA, 101142229). Views and opinions expressed are however those of the author(s) only and do not necessarily reflect those of the European Union or the European Research Council Executive Agency. Neither the European Union nor the granting authority can be held responsible for them. For the purpose of Open Access the author has applied a CC BY public copyright licence to any Author Accepted Manuscript version arising from this submission. We thank Marc Ponsen from Statistics Netherlands for valuable input on various aspects of the study.
\end{acks}

\bibliographystyle{ACM-Reference-Format}
\bibliography{refs}



\end{document}